\documentclass[aps,prl,showpacs,twocolumn,superscriptaddress]{revtex4}
\usepackage{graphicx}
\usepackage{amsmath}
\usepackage{color}
\def\etal{$\it{et~al.}$}

\begin{document}

\title{Infrared spectroscopy of Landau levels in graphene}

\author{Z. Jiang}
\altaffiliation{Electronic address: jiang@magnet.fsu.edu}
\affiliation{Department of Physics, Columbia University, New York, New York 10027, USA}
\affiliation{National High Magnetic Field Laboratory, Tallahassee, Florida 32310, USA}

\author{E. A. Henriksen}
\affiliation{Department of Physics, Columbia University, New York, New York 10027, USA}

\author{L. C. Tung}
\affiliation{National High Magnetic Field Laboratory, Tallahassee, Florida 32310, USA}

\author{Y.-J. Wang}
\affiliation{National High Magnetic Field Laboratory, Tallahassee, Florida 32310, USA}

\author{M. E. Schwartz}
\affiliation{Department of Physics, Columbia University, New York, New York 10027, USA}

\author{M. Y. Han}
\affiliation{Department of Applied Physics and Applied Mathematics, Columbia University, New York, New York 10027, USA}

\author{P. Kim}
\affiliation{Department of Physics, Columbia University, New York, New York 10027, USA}

\author{H. L. Stormer}
\affiliation{Department of Physics, Columbia University, New York, New York 10027, USA}
\affiliation{Department of Applied Physics and Applied Mathematics, Columbia University, New York, New York 10027, USA}
\affiliation{Bell Labs, Lucent Technologies, Murray Hill, New Jersey 07974, USA}
\date{\today}

\begin{abstract}
We report infrared studies of the Landau level (LL) transitions in single layer graphene. Our specimens are density tunable and show \textit{in situ} half-integer quantum Hall plateaus. Infrared transmission is measured in magnetic fields up to $B=18$ T at selected LL fillings. Resonances between hole LLs and electron LLs, as well as resonances between hole and electron LLs are resolved. Their transition energies are proportional to $\sqrt{B}$ and the deduced band velocity is $\tilde{c}\approx1.1\times10^6$ m/s. The lack of precise scaling between different LL transitions indicates considerable contributions of many-particle effects to the infrared transition energies.
\end{abstract}

\maketitle

Graphene is the newest member in the family of two-dimensional (2D) carrier systems, which have shown a spectrum of fascinating new physics over the past decades. Graphene, a single atomic sheet of graphite, represents the ultimate 2D material. Moreover, its electronic band structure differs radically from the parabolic bands common to all previous 2D systems. In graphene, the conduction and valence bands meet at two inequivalent, charge-neutral points in momentum space around which the dispersion is linear, leading to the so-called Dirac cones \cite{mcclure56}. Much of the interest in graphene stems from an analogy of this dispersion relation to that of relativistic, massless fermions, leading to intriguing new phenomena. For instance, a very unusual half-integer quantum Hall effect (QHE) and a non-zero Berry's phase have been discovered in graphene \cite{novoselov05-1,zhang05,zhang06}, as well as an abnormally weak localization \cite{morozov06,wu06}. More recently, Raman spectroscopy \cite{ferrari06,draf06,yan06,pisana06} and angle resolved photoelectron spectroscopy \cite{bostwick06} have been applied to graphene, yielding information on electron-phonon coupling and on the energy dispersion of the Dirac cones. 

Infrared (IR) spectroscopy is a powerful tool for investigating the low-lying energy excitations of a material. When combined with a magnetic field, $B$, it allows for the study of its Landau level (LL) spectrum. In traditional 2D materials with parabolic dispersions this is tantamount to measuring the carrier effective mass, $m^*$, since transitions between the equally spaced LLs at energy $E_n=(n+1/2)\hbar eB/m^*$ reflect the same $m^*$ as in classical cyclotron resonance at $\omega_c=eB/m^*$. Here $e$ is the electron charge, $\hbar$ is Planck's constant, and the non-negative integer $n$ is the LL index. 

In a magnetic field, the linear dispersion relation of graphene leads to an unequally spaced LL spectrum \cite{mcclure56,haldane88,semenoff84},
\begin{equation}
E_n=\text{sgn}(n)\times\sqrt{2e\hbar\tilde{c}^2B\left|n\right|}=\text{sgn}(n)\times\sqrt{2\left|n\right|}\hbar\tilde{c}/l_0
\end{equation}
where $\tilde{c}$ is the band velocity, $l_0=\sqrt{\hbar/eB}$ is the magnetic length, and $n>0$ or $n<0$ represents electrons or holes, respectively. Most unusually, for $n=0$ there exists a LL at $E_0=0$ with a distinctive electron-hole degeneracy. This peculiar behavior of carriers in graphene is further enriched by spin splittings \cite{zhang06}, possible lifting of the Dirac cone valley degeneracy and general many-particle effects, and provides a unique opportunity to probe these exceptional electronic properties of graphene via LL formation.

In this letter, we report IR transmission results on single layer graphene. In fields up to $B=18$ T, two identifiable LL transitions are clearly resolved, and their energy position scales as $\sqrt{B}$ with a slope corresponding to a $\tilde{c}\approx1.1\times10^6$ m/s in Eq. (1). A deviation from an ideal ratio of $1:(\sqrt{2}+1)$ between the two transition energies, predicted by simple matrix elements \cite{gusynin06}, indicates a considerable many-particle contribution \cite{iyengar06} to these LL transitions. 

Sadowski \textit{et al} reported LL spectroscopy of ultrathin graphite layers created by thermal deposition on SiC \cite{sadowski06}. Their data - at considerably lower magnetic field - also show $\sqrt{B}$ behavior and they deduce a $\tilde{c}$ similar to ours. However, the nature of the thin graphite sheet remains unclear. 

Our studies were performed on graphene, mechanically cleaved from bulk Kish graphite and deposited onto lightly doped Si/SiO$_2$ substrates \cite{novoselov05-2} which are transparent to IR, yet sufficiently conductive to serve as gates. The primary challenge of the experiment is the mismatch between the size of the IR focus in typical light-pipe systems ($\sim$1 mm) and the lateral dimension of typical mechanically extracted graphene samples ($\sim$10 $\mu$m). To overcome this limitation, we selected only large graphene specimens with areas as big as a few thousand $\mu$m$^2$ and employed a parabolic cone to focus the IR light to a few hundred $\mu$m spot. Standard e-beam lithography, metal evaporation and lift-off techniques were used to define the Cr/Au (3/35 nm) contact wires, along with a $\sim$100 $\mu$m diameter metal aperture around the specimen, which reduces stray IR light around the graphene sample, see inset to Fig. 1. The graphene device was mounted on the parabolic cone so the IR light passes through the Si substrate onto the graphene. The substrate was thinned and wedged to 4$^\circ$ to suppress Fabry-Perot interferences and could be aligned \textit{in situ} with respect to the IR focus. A composite Si bolometer directly beneath the sample served as the IR detector. All electrical and IR experiments were performed at 4.2 K in magnetic fields up to $B=18$ T using conventional quasi-dc lock-in technique and a Bruker IFS 66v/S FTIR spectrometer. 

\begin{figure}[t]
\includegraphics[width=8cm]{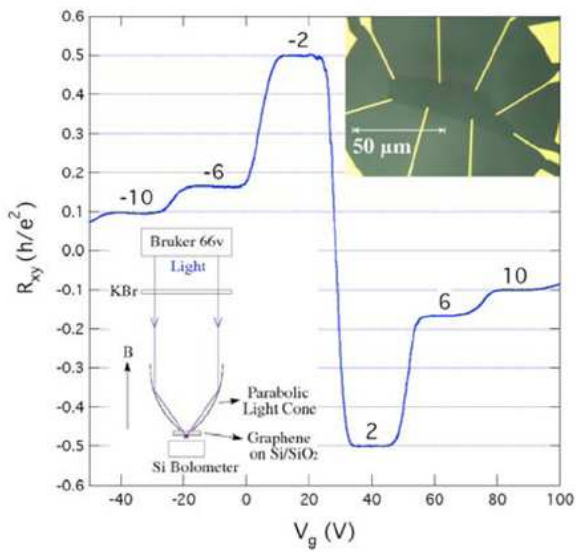}
\caption{(color online). Half-integer quantum Hall effect in graphene with plateaus at $R_{xy}=h/\nu e^2$ for filling factors $\nu=\pm2, \pm6, \pm10$, at 4.2 K and 18 T. LL filling factors are indicated. Note that our devices are density tunable with $n_s/V_g=7.2\times10^{10}$ cm$^{-2}$V$^{-1}$, where $n_s$ is the charge carrier density in graphene. Upper inset: optical image of an 1100 $\mu$m$^2$ graphene device. Lower inset: schematic of the experimental setup.}
\end{figure}

The devices used in our work exhibit mobilities of $2-4\times10^3$ cm$^2$V$^{-1}$s$^{-1}$ measured at densities of $\sim2\times10^{12}$ cm$^{-2}$. The gate voltage corresponding to charge neutrality, $V_{Dirac}$, is $\sim$29 V due to a built-in potential from residual charges in the environment. Figure 1 shows a typical Hall resistance trace ($R_{xy}$) vs. gate voltage for fixed $B=18$ T taken \textit{in situ}. The characteristic half-integer QHE with plateaus at $R_{xy}=h/\nu e^2$ for filling factors $\nu=\pm2, \pm6, \pm10$ \cite{novoselov05-1,zhang05} is clearly observed, confirming the single layer nature of our graphene specimen.

We record IR transmission spectra at fixed $B$ field at two different carrier concentrations, corresponding to two different integer LL fillings. In this way the experimental conditions are identical for all system components, such as bolometer sensitivity, silicon transparency and $B$-induced shifts of optical path. The only change is in the LL occupation of the graphene sample. Furthermore, since the LL spacing - and hence the IR resonance - is density (LL filling factor) dependent, the two spectra, although at identical $B$-field, show transmission minima at two different IR frequencies and thus their ratio results in a maximum and a minimum placed on a background of 1. Figure 2 shows select transmission spectra of the sample shown in Fig. 1 at different magnetic fields following this normalization method, for two traces at filling factors $\nu=-2$ (numerator) and $\nu=-10$ (denominator). Given the fourfold LL degeneracy of graphene and the electron-hole symmetry of the $n=0$ LL, filling factors $\nu=-2$ and $\nu=-10$ correspond to a Fermi level ($E_F$) position between the $n=-1$ and $n=0$ LLs and the $n=-3$ and $n=-2$ LLs, respectively; see inset to Fig. 2. At $B=18$ T, switching the gate voltages between these two positions tunes the hole density from $9.4\times10^{11}$ cm$^{-2}$ to $4.7\times10^{12}$ cm$^{-2}$.

\begin{figure}[t]
\includegraphics[width=8cm]{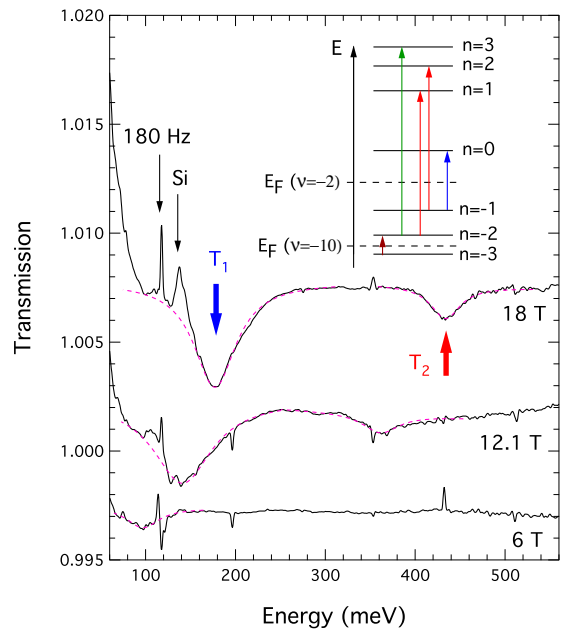}
\caption{(color online). Normalized IR absorption spectra of holes in graphene at three different magnetic fields, taken by dividing spectra taken at filling factors $\nu=-2$ and $\nu=-10$. Two LL resonances are denoted by T$_1$ and T$_2$. Residual spectral artifacts are associated with 60 Hz harmonics and with carriers in the Si substrate. Dashed purple lines are Lorentzian fits to the data. The inset shows a schematic LL ladder with allowed transitions indicated by arrows.}
\end{figure}

Two transmission minima, T$_1$ and T$_2$, are readily observable in the traces of Fig. 2. Their minima associate them with the $E_F(\nu=-2)$ position. Features from the $E_F(\nu=-10)$ position occur as maxima residing at lower energies, and are only visible as shoulders in the 18T and 12.1T traces. Therefore, in Fig. 2 the $E_F(\nu=-10)$ spectra simply serve as a normalization for the $E_F(\nu=-2)$ spectra.

Remaining artifacts in the data are largely due to 60 Hz harmonics (narrow spikes). In addition, a feature associated with the Si substrate emerges, since the introduction of carriers in graphene leads to the same density but opposite sign of carriers at the Si-SiO$_2$ interface \cite{basov07}. The transmission of this carrier system is density dependent and therefore does not completely average out in spectral division. However, these artifacts in Fig. 2 can be circumvented and we find that the minima can be fit well by a Lorentzian, shown by the dashed lines. From these fits we determine the resonance position and the half-width of each resonance. 

The transmission minima in Fig. 2 are well developed and their resonance energies clearly decrease with decreasing magnetic field. Evidently, T$_2$ is considerably weaker than T$_1$ and the ratio of their intensities seems to be roughly constant. The widths of the resonances, $\delta E$, are similar for T$_1$ and T$_2$ and correspond to a scattering time, $\tau\cong\hbar/\delta E\cong20$ fs in reasonable agreement with $\tau\cong40$ fs, derived from the mobility. 

\begin{figure}[b]
\includegraphics[width=8cm]{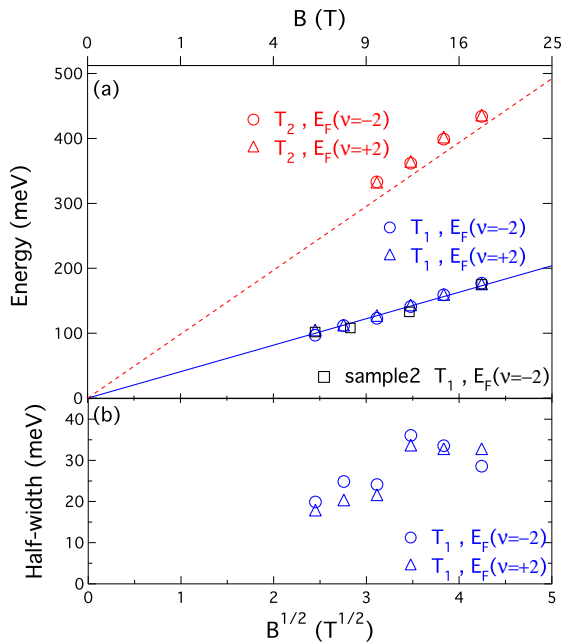}
\caption{(color online). (a) Resonance energies vs. $\sqrt{B}$, from holes (ratio of $\nu=-2$ and $\nu=-10$ data, Fig. 2) and electrons (ratio of $\nu=+2$ and $\nu=+10$, spectra not shown). Similar behavior has been observed in a second sample (squares). Solid line is a best $\sqrt{B}$-fit to the T$_1$ transition, yielding $\tilde{c}=(1.12\pm0.02)\times10^6$ m/s. The dashed line represents a scaling of the solid line by a factor $(\sqrt{2}+1)$. (b) Half-width at half maximum from Lorentzian fits of the T$_1$ transitions, as a function of $\sqrt{B}$.}
\end{figure}

Figure 3(a) summarizes the energies of the T$_1$ and T$_2$ transitions determined from Lorentzian fits to spectra from holes (Fig. 2) and electrons (not shown), plotted as a function of $\sqrt{B}$. A clear $\sqrt{B}$ relationship emerges. This dependence is very reliable for the T$_1$ transition with a large number of data points to low $B$-field. According to Eq. (1) a $\sqrt{B}$ dependence is expected for any transition between LLs in graphene and our data confirm this. Inspecting the slope and comparing it with Eq. (1) assuming $\tilde{c}\approx1\times10^6$ m/s \cite{novoselov05-1,zhang05,dresselhaus02,brandt88} identifies the transition as the $n=-1 \rightarrow n=0$ (holes) and $n=0 \rightarrow n=1$ (electrons) intraband LL transitions. Quantitatively, this assignment leads to a refined value of $\tilde{c}=(1.12\pm0.02)\times10^6$ m/s, indicated as a line through the data in Fig. 3(a). Given the Fermi level position $E_F(\nu=-2)$ shown in the inset to Fig. 2 these are the lowest LL transitions possible in graphene. Fig. 3(b) shows the $B$-dependence of the half-width at half maximum of transition T$_1$. It seems to be increasing, but its $B$-dependence cannot be further assessed at this stage.

Had this been a traditional 2D electron or hole system we would have observed only one resonance line, the cyclotron resonance at $\hbar\omega_c=\hbar eB/m^*$, from which arises a unique mass, $m^*$. In contrast, we observe a second, higher energy transition, T$_2$. To identify the LLs involved requires a recapitulation of the relevant dipole transition selection rules. In a traditional 2D system with energy spectrum $E_n=(n+1/2)\hbar\omega_c$ they dictate $\Delta n=n_2-n_1=\pm1$ for either electron or hole Landau ladders. When expanded to semiconductors in which interband transitions from a hole LL, $n_h$, to an electron LL, $n_e$, are also feasible, the selection rule changes to $\Delta n=n_e-n_h=0$ due to the separate \textit{p} and \textit{s} character of valence and conduction band. In graphene, in which ``valence'' and ``conduction band'' have the same symmetry, a single selection rule $\Delta n=\left|n_2\right|-\left|n_1\right|=\pm1$ is obtained \cite{gusynin06} for intra as well as for interband transitions. Therefore, the next allowed LL transitions for $E_F(\nu=-2)$ with energy above the $n=-1 \rightarrow n=0$ transition are interband transitions $n=-2 \rightarrow n=1$ and $n=-1 \rightarrow n=2$, see inset Fig. 2. While preserving the general $\sqrt{B}$ dependence their energy scales as $1:(\sqrt{2}+1)$ compared to the $n=-1 \rightarrow n=0$ intraband transition, following Eq. (1). The dashed line in Fig. 3(a) represents a scaling of the T$_1$ transition by a factor of $\sqrt{2}+1$ and falls quite close to the T$_2$ data. This provides very good evidence that T$_2$ represents the lowest interband transition expected from the inset of Fig. 2 and its electron-hole symmetric equivalent. 

The observed small deviation of the $1:(\sqrt{2}+1)$ ratio of the T$_1$ and T$_2$ energies, however, lies well outside of the experimental errors ($\sim$symbol size). A $\sqrt{B}$-fit to T$_2$ with a zero energy intercept results in $\tilde{c}=(1.18\pm0.02)\times10^6$ m/s, according to Eq. (1). This value differs appreciably from the value $\tilde{c}=(1.12\pm0.02)\times10^6$ m/s deduced from T$_1$, as well as from the value of $\tilde{c}=(1.03\pm0.01)\times10^6$ m/s found by Sadowski \textit{et al} \cite{sadowski06}, who derive the same band velocity $\tilde{c}$ for both transitions. 

The discrepancy between the band velocities deduced from different LL transitions sheds doubt on the applicability of a simple LL energy subtraction scheme based on Eq. (1) for the interpretation of IR data in graphene. In 2D systems with parabolic dispersions, Kohn's theorem \cite{kohn61} explains that e-e interactions have no impact on the LL transition energies observed in IR experiments (cyclotron resonance). Instead, the resonance energy coincides with the non-interacting value, provided that the system is translationally invariant, which, in spite of residual disorder, can largely be assumed to hold. Yet, Kohn's theorem fails in the case of graphene, whose linear dispersion may be viewed as a case of extreme non-parabolicity, and thus many-particle effects may be expected to contribute to the LL transition energies. 

Indeed, the first calculations of many-particle corrections to the bare LL transitions are appearing in the literature. A recent paper by Iyengar \textit{et al} \cite{iyengar06} arrives at quite large many-particle contributions, on the order of a few $e^2/\epsilon l_0=56\ \text{meV}\ \sqrt{B\text{(T)}}/\epsilon$, which amounts to an energy of $\sim$60 meV at $B=18$ T with an assumed dielectric constant of $\epsilon=4$. Importantly, the total transition energy between LLs with indices $n$ and $m$ scales as $\Delta E_{n,m}\times l_0=\sqrt{2}\times\hbar\tilde{c}(\sqrt{\left|m\right|}\pm\sqrt{\left|n\right|})+C_{n,m}\times e^2/\epsilon$, so that contributions from $\tilde{c}$ and $C_{n,m}/\epsilon$ cannot be distinguished by their $B$-dependence.

On the other hand, the magnitude of the many-particle effects, $C_{n,m}$, depends on the specific $n$, $m$ LL pair, a fact that can be employed to distinguish them from the band velocity contribution $\tilde{c}$. Iyengar \textit{et al} \cite{iyengar06,wang07} calculated the prefactors $C_{-1,0}=1.18$ and $C_{-2,1}=C_{-1,2}=3.17$ for the transitions T$_1$ and T$_2$ of Fig. 3(a), respectively. Since $C_{-2,1}/C_{-1,0}>(\sqrt{2}+1)$, any e-e contribution to the LL transition will increase the ratio of transition energies beyond its non-interacting value of $\sqrt{2}+1$, as is observed in our data. In fact, a band velocity $\tilde{c}\approx8\times10^5$ m/s and $\epsilon\approx5$ can fit simultaneously the T$_1$ and T$_2$ data of Fig. 3(a). However, at this stage a wide range of $\tilde{c}$ and $\epsilon$ produce similar good fits within the error bars of the experiment making it premature to assign any particular values. More precise IR measurements between several other, higher LLs are required to assess quantitatively the impact of many-particle physics on these transitions. However, it appears likely at this stage that a determination of the band velocity from a single particle LL picture leads to a considerable overestimation of its value. A quantitative comparison of our data with many-body calculations \cite{iyengar06,wang07} may be premature, since theory does not take into account disorders \cite{peres06,dora07} nor any mesoscopic corrugations of the graphene sheet \cite{morozov06}.

Finally, apart from many-particle effects, one may wonder as to the impact of a possible gap opening around the Dirac point due to interaction effects, a scenario that has been addressed by Gusynin \textit{et al} \cite{gusynin06}. In this model, any positive value for the gap energy reduces the ratio of T$_2$ to T$_1$ below $\sqrt{2}+1$, in contrast to our data.

In summary, we have observed intra- and inter-LL transitions in IR spectroscopy on graphene. The transition energies scale as $\sqrt{B}$ and a simple, non-interacting LL transition interpretation yields a band velocity of $\tilde{c}\approx1.1\times10^6$ m/s. The observed deviation from a precise $1:(\sqrt{2}+1)$ scaling between the transition energies indicates a contribution from many-particle interactions to the transitions. Theory predicts rather strong ($\sim$$30\%$) such corrections \cite{iyengar06,wang07}. Qualitatively, the observed deviations show the correct sign, but the magnitude of the corrections cannot yet be deduced reliably from experiment. 

We would like to thank Kun Yang, H.A. Fertig, L. Brey, Jianhui Wang, and D.N. Basov for discussions and B. Oezyilmaz for assistance in fabrication. This work is supported by the DOE (DE-AIO2-04ER46133 and DE-FG02-05ER46215), NSF (DMR-03-52738 and CHE-0117752), NYSTAR, and the Keck Foundation. IR measurement of this work was performed at the National High Magnetic Field Laboratory, which is supported by NSF Cooperative Agreement No. DMR-0084173, by the State of Florida, and by the DOE. We thank J. Jaroszynski, E.C. Palm, T.P. Murphy, S.W. Tozer, and B.L. Brandt for experimental assistance. Z. Jiang and E.A. Henriksen contributed equally to this work.

\end{document}